\documentclass{Interspeech}

\interspeechcameraready

\usepackage{multirow}
\usepackage{tablefootnote}
\usepackage{xcolor}
\usepackage{adjustbox}
\usepackage{booktabs}
\usepackage{makecell}

\title{Distilling a speech and music encoder with task arithmetic}
\author[affiliation={1,2}]{Fabian}{Ritter-Gutierrez$^{*}$}
\author[affiliation={3}]{Yi-Cheng}{Lin$^{*}$}
\author[affiliation={3}]{Jui-Chiang}{Wei}
\author[affiliation={2}]{Jeremy H.M}{Wong}
\author[affiliation={1}]{Eng-Siong}{Chng}
\author[affiliation={2}]{Nancy F.}{Chen}
\author[affiliation={3}]{Hung-yi}{Lee}

\affiliation{}{Nanyang Technological University}{Singapore} \affiliation{}{Institute for Infocomm Research (I2R)}{Singapore}
\affiliation{}{National Taiwan University}{Taiwan}

\email{s220064@e.ntu.edu.sg, f12942075@ntu.edu.tw}
\keywords{self-supervised models, knowledge distillation}

\usepackage{comment}

\begin{document}

\maketitle

\begingroup
  \renewcommand\thefootnote{*}
  \footnotetext{Equal contribution}
\endgroup

\begin{abstract}
Despite the progress in self-supervised learning (SSL) for speech and music, existing models treat these domains separately, limiting their capacity for unified audio understanding. A unified model is desirable for applications that require general representations, e.g. audio large language models. Nonetheless, directly training a general model for speech and music is computationally expensive. Knowledge Distillation of teacher ensembles may be a natural solution, but we posit that decoupling the distillation of the speech and music SSL models allows for more flexibility. Thus, we propose to learn distilled task vectors and then linearly interpolate them to form a unified speech+music model. This strategy enables flexible domain emphasis through adjustable weights and is also simpler to train. Experiments on speech and music benchmarks demonstrate that our method yields superior overall performance compared to ensemble distillation.



\end{abstract}

\section{Introduction}\label{sec:introduction}

Speech/Music self-supervised learning models (SSL) have significantly advanced speech and music processing tasks. Notably, HuBERT \cite{hubert} has demonstrated exceptional performance in speech representation learning, while MERT \cite{mert} has been tailored for music understanding. These models have shown that large-scale pre-training can effectively capture domain-specific features. However, current SSL models treat speech and music as separate domains. 

While separate SSL models are useful for domain-specific downstream task evaluations, having instead a single general model, able to capture the complexities of both speech and music decreases computational memory requirements by eliminating the decision of choosing a model depending on the audio input. This is particularly beneficial for audio large language models (ALLMs) \cite{salmon, dynamicsuperbv2}. ALLMs rely on their encoders to interpret complex audio inputs. For example, in \cite{salmon}, the embeddings from both speech and audio encoders are concatenated together and used for the LLM. However, such an approach has a computational cost that scales with the number of encoders. Hence, a more general representation can benefit ALLMs.


An intuitive approach to get a unified speech+music representation would be to pre-train a single model on both domains, perhaps using a combination of SSL losses such as the ones in HuBERT and MERT. However, pre-training from scratch is computationally expensive and most times infeasible for academic research \cite{liu2024efficient}. A more practical alternative is knowledge distillation \cite{jinyu_distill,hintonDistill}, where a smaller model can learn from multiple larger teacher models \cite{jeremy-multi-distill,ensemble_distillation_speech}, one specialized in speech and another in music, thus integrating knowledge from both domains while reducing the overall model size. Nonetheless, certain problems arise under this strategy. 

One, when learning from multiple teachers that specialize in different domains, conflicting domain-specific signals may arise, thus hurting learning of a unified representation. Second, GPU requirements for distillation pre-training grow with each additional teacher, thus compounding the same computational barriers that make large-scale pre-training infeasible for many academic labs. Crucially, this approach lacks flexibility; adjusting the emphasis between speech and music requires complete retraining.  This raises the key question: how can we achieve efficient, conflict-minimized, and flexible distillation for a unified speech+music model?.

Model merging is a research direction that interpolates parameters from different models to create a single model that inherits the capabilities of its constituent ones \cite{permutation_linear_interpolation,task_arithmetic}. In particular, the work on ``task arithmetic" \cite{task_arithmetic} demonstrates that one can obtain task-specific “vectors” by fine-tuning a pre-trained model on different specialized tasks. These task vectors can then be added to the original pre-trained model to effectively incorporate the specialized knowledge from each task. Existing task arithmetic work \cite{task_arithmetic_synthetic_speech,task_arithmetic_tts, task_arithmetic_speech_translation}, however, focuses solely on supervised fine-tuning. We extend the concept of task arithmetic to distillation pre-training.

\begin{figure*}[!th]
  \centering
  \includegraphics[width=0.93\linewidth]{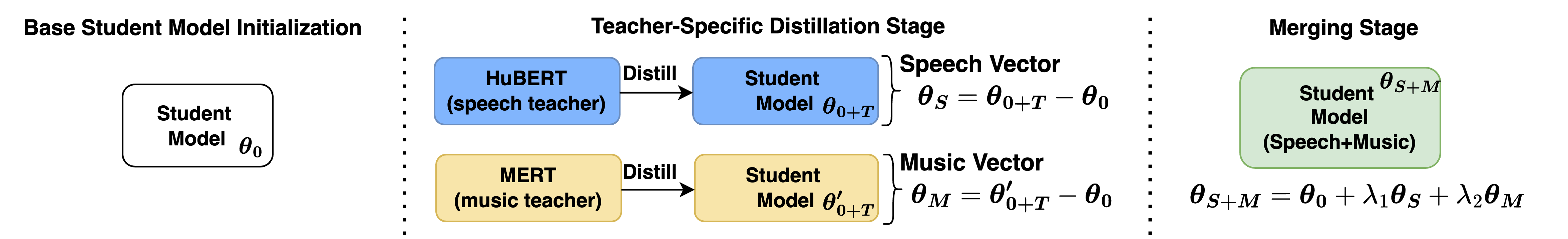}
  \vspace{-3mm}
  \caption{Overview of the distillation method via task arithmetic.}
  \label{fig:proposal_task_arithmetic}
  \vspace{-4mm}
\end{figure*}

This paper proposes to combine the benefits of distillation and task arithmetic model merging for the distillation of a speech+music SSL model. The method, as shown in Fig \ref{fig:proposal_task_arithmetic}, separately distills each of the teacher models into two separate student models. A speech and music vector can be computed as the arithmetic difference of the distilled model weights with its initialized weights. Then, a speech+music model can be constructed via linear interpolation of the speech and music vectors. 

This design not only mitigates the interference encountered in ensemble of teachers distillation but also reduces GPU memory requirements by avoiding the simultaneous processing of multiple teacher models. Moreover, the method offers enhanced flexibility: by simply adjusting the interpolation weights, one can emphasize speech or music representations on demand. In contrast, an ensemble distillation approach would need full retraining from scratch whenever a new balance between speech and music is desired, thus limiting adaptability. 

Results on standard speech and music tasks show that task arithmetic can yield a better model compared to distilling both teachers simultaneously. Ablations studies are provided to show the performance effect on the speech and music tasks depending on the interpolation weights chosen. In general, the more speech weight, then the better performance on speech tasks and same otherwise.

\vspace{-0.1pt}

\section{Related Work}
\textbf{Knowledge distillation:} Knowledge distillation \cite{jinyu_distill,hintonDistill} is a model compression method that creates a small student model which learns to behave like a larger teacher model. In the context of speech self-supervised learning (SSL), several works have focused on distilling speech representations. DistilHuBERT \cite{distilhubert} distills the knowledge of HuBERT into a smaller architecture, showing competitive performance on speech processing tasks while reducing model size. Similarly, FitHuBERT \cite{fithubert} and LightHuBERT \cite{lighthubert} explored distillation strategies to optimize the trade-off between model size and performance. Other works \cite{RobustDistilHuBERT,robustdistiller,ritter2023noise} explores noise robustness of distilled SSL representations or sequence level compression of DistilHuBERT \cite{meng2023compressing,onceforallcompression}. In the music domain, there has been no direct work on distilling music SSL representations. However, previous studies have explored knowledge distillation in audio model architectures. For example, \cite{gao2022multi}, learns different audio features transformations. Each of such features are aggregated into a single model via distillation.

However, such methods distills only single-domain models, leaving unaddressed the challenges of unifying knowledge from multiple domains. Some works have explored distilling ensembles of teachers in the context of speech representation learning. For instance, \cite{ensemble_distillation_speech} distills from multiple speech SSL models by adding extra prediction heads on a distilHuBERT architecture. While effective, such an approach adds extra computational hurdles.

\noindent\textbf{Task arithmetic:}
Task arithmetic adapts pre-trained models to multiple tasks by first fine-tuning on a specific task \cite{task_arithmetic}, then computing a task vector as the difference in model parameters before and after fine-tuning.  Knowledge can be combined across tasks by linearly interpolating different task vectors. This approach has been successfully applied to speech domain, such as domain adaptation in automatic speech recognition (ASR) \cite{task_arithmetic_synthetic_speech}, text-to-speech synthesis \cite{task_arithmetic_tts}, and speech translation \cite{task_arithmetic_speech_translation}.

While task arithmetic has demonstrated effectiveness in adapting pre-trained models for various tasks \cite{task_arithmetic,model_soup}, its application has primarily been within supervised learning paradigms.  However, in contrast to these supervised approaches, the methodology proposed in this paper focuses on leveraging task arithmetic to unify representations from distinct SSL domains. Our objective is not to combine tasks through supervised fine-tuning of a single pre-trained model, but rather to explore the potential of task arithmetic for directly merging knowledge encoded in independently pre-trained SSL models for speech and music.


\vspace{-2.5mm}
\section{Task Arithmetic for knowledge distillation}
\vspace{-1mm}

Generally speaking, task arithmetic is defined by task vectors under supervised fine-tuning. Namely, let $\boldsymbol{\theta}^{task}$ denotes a task vector with a dimensionality composing the whole neural architecture model parameters. Such vector is obtained by computing the model parameters difference from a pre-trained model with parameters $\boldsymbol{\theta}^{PT}$ and a fully fine-tuned model parameters, $\boldsymbol{\theta}^{FT}$, trained in a particular task, , e.g. speech translation. Concretely, $\boldsymbol{\theta}^{task} = \boldsymbol{\theta}^{FT} - \boldsymbol{\theta}^{PT}$.

This paper differs from the supervised scenario by casting the problem for distillation. We note that task arithmetic can be applied for knowledge distillation as long as each distilled model has the same student weight parameter initialization. Concretely, let $\boldsymbol{\theta}_0$ denotes the model parameter initialization of the student model. Here, the goal is to distill two teacher models from different domains, one for speech and another for music. Rather than employing an ensemble model distillation \cite{ensemble_distillation_speech} which requires more GPU memory, we distill each model separately. Particularly, let $\boldsymbol{\theta}_{S} = \boldsymbol{\theta}_{0+T} - \boldsymbol{\theta}_{0}$ be the speech vector, which is trained by distilling from HuBERT teacher on $T$ iterations and let $\boldsymbol{\theta}_{M} = \boldsymbol{\theta}^{'}_{0+T} - \boldsymbol{\theta}_{0}$, be the music vector, which is trained under the same number of iterations but the teacher model changes to MERT. Then, a speech+music model can be created by the linear interpolation,

\begin{equation}
\boldsymbol{ \theta}_{S+M}=\boldsymbol{\theta}_0+\lambda_1 \boldsymbol{\theta}_S+\lambda_2 \boldsymbol{\theta}_M
\end{equation}

where $\lambda_1$ and $\lambda_2$ are manually chosen interpolation weights that allow flexible adjustment of the model’s focus toward either speech or music, generally $\lambda_1 + \lambda_2 = 1$. This interpolation framework provides significant advantage over the ensemble distillation approach \cite{ensemble_distillation_speech}, which requires simultaneous optimization over multiple domains which may lead to student-learning interference. Our method avoids potential interference between tasks during distillation and reduces GPU memory overhead by decoupling the learning processes for each task.

\vspace{-2mm}
\section{Experimental setup}

\subsection{Knowledge Distillation}
\vspace{-0.5mm}
Distillation is done by using LibriSpeech 960 dataset \cite{librispeech} and the student model architecture, and the distillation loss function is the same as DistilHuBERT \cite{distilhubert}. LibriSpeech 960 was selected as the primary distillation dataset because preliminary experiments indicated that the teacher signal was more influential on student model performance than the specific domain of the distillation data itself (See Table \ref{tab:pretrain_data_table}). Two different teacher models are distilled, one corresponds to HuBERT Base\footnote{https://huggingface.co/facebook/hubert-base-ls960} \cite{hubert}, i.e., HuBERT with 12 transformer layers as well as MERT Base\footnote{https://huggingface.co/m-a-p/MERT-v0-public} \cite{mert} (mert-v0-public) which has as well 12 transformer layers. HuBERT is tailored for speech related task as evaluated in SUPERB \cite{superb} and it was pre-trained with LibriSpeech 960 as well. On the other hand, MERT (mert-v0-public) is tailored for music understanding tasks and it was pre-trained with Music4All dataset \cite{music4all}. For single teacher model distillation, we adopt the multi-layer distillation strategy from DistilHuBERT \cite{distilhubert}, employing three prediction heads attached to the 4th, 8th, and 12th transformer layers of the teacher model. This approach is maintained regardless of the teacher model. The student model is initialized by copying the CNN encoder and first and second transformer layers from HuBERT when distilling either of the teachers. Using the same starting point for both student models ensures that the resulting parameter differences are directly comparable, thereby enabling meaningful linear interpolation when merging the speech and music task vectors. Initializing from HuBERT was chosen because preliminary empirical evaluations indicated that HuBERT-initialized student models yielded superior performance in initial downstream task assessments compared to alternative initializations.

For ensemble distillation, following \cite{ensemble_distillation_speech}, we employed separate prediction heads for each teacher model (HuBERT and MERT). This resulted in a student model with six prediction heads, allowing simultaneous learning from layers 4, 8, and 12 of both teacher networks (see Eq. 6 in \cite{ensemble_distillation_speech}). Consistent with single-teacher distillation, the student model was also initialized from HuBERT. All experiments are run for 25 epochs (200k steps), and we use the final checkpoint for downstream evaluations to compare the performance of each model under the same pre-training optimization trajectory.

\vspace{-2.5mm}

\subsection{Downstream Tasks Evaluation}

We choose a mix of speech and music tasks to evaluate all the models. For the speech tasks, we select the Automatic Speech Recognition (ASR), Keyword Spotting (KS), Intent Classification (IC), Emotion Recognition (ER) and Speaker Identification (SID) from the SUPERB Benchmark \cite{superb}. Such task selection allows us to encompass the semantics, speaker, and paralinguistic capability of the models to be evaluated. We use the s3prl toolkit to assess performance. Additionally, we implement into s3prl some music tasks derived from MARBLE Benchmark \cite{marble}. In particular the tasks Singer Identification (SingerID) and Pitch Classification (PitchID).  SUPERB tasks were kept as it is while for the MARBLE task, we implemented it by following the same simple downstream linear layers principle used in KS, IC, ER, and SID. Additionally, both SingerID and PitchID use Adam optimizer with 1E-3 learning rate and a batch size of 64. This downstream architecture selection closely follows the one of MARBLE (constrained task).

Additionally, to evaluate overall performance across all the tasks, we follow two ranking methods. The first is motivated by the SUPERB Score defined in Eq. (1) in \cite{mlsuperb}, where rather than using the SOTA value for the denominator, we use the best performance achieved between HuBERT and MERT. SUPERB score assesses representation ability across all task domains and the higher the value, the better. The second ranking method is an average ranking across each task. This means that each task is ranked in terms of best performance, e.g. HuBERT ranks 1 for ASR while Task arithmetic ranks 3. Then each rank is averaged to get a Rank average metric. Here, the lower, the better. 

\vspace{-1mm}
\section{Experiments}\label{sec:Experiments}
\textbf{Question 1:} \textit{Why can't we directly merge HuBERT and MERT rather than distill them and merge them?}.\\

\noindent Table~\ref{tab:main_results} demonstrates that naively merging HuBERT and MERT by averaging their weights (denoted as \textit{(HB + MR) / 2}) leads to poor performance across all tasks. For instance, the merged model achieves a WER of 37.44\% on ASR (compared to HuBERT’s 6.42\%) and a SingerID accuracy of 52.17\% (compared to MERT’s 85.69\%). This collapse in performance suggests that HuBERT and MERT, despite sharing a similar transformer-based architecture, occupy distinct regions in the weight space optimized for their respective domains (speech and music). Direct interpolation of their weights places the merged model in a suboptimal region that fails to retain meaningful representations for either domain. This aligns with theoretical observations in model merging literature \cite{permutation_linear_interpolation}, which posit that merging models from divergent basins of the loss landscape leads to destructive interference. On the other hand, distilling both HuBERT and MERT into student models initialized from shared weights ensures that the distilled models (distilHuBERT and distilMERT) lie in a closer region of the weight space. This alignment is critical for enabling task arithmetic, as shown in prior work \cite{linear_inter_lottery_hyp,permutation_linear_interpolation}, where shared initialization facilitates parameter interpolation.

\noindent \textbf{Question 2:} \textit{Is is possible to learn a speech+music distilled model with task arithmetic?}.\\

\noindent Our results show that task arithmetic achieves superior performance over ensemble distillation across most tasks while requiring significantly less computational overhead. As shown in Table~\ref{tab:main_results}, task arithmetic outperforms ensemble distillation on ASR (13.97\% vs. 15.04\% WER), KS (95.98\% vs. 94.29\% accuracy), and SingerID (83.35\% vs. 82.21\% accuracy). Notably, task arithmetic achieves a higher SUPERB score (801.78 vs. 800.47) and a better average rank (3.00 vs. 3.71), indicating stronger generalization across both domains. While the SUPERB Score and rank are quite similar between both approaches, such results indicates that using task arithmetic for distillation is indeed possible and future work on how to improve the overall performance across all the tasks with model merging is a worth exploring direction.

We identify two key advantages of task arithmetic:
\begin{enumerate}
    \item \textbf{Computational Efficiency:}  Unlike ensemble distillation, which requires training with six prediction heads (to learn from both HuBERT and MERT layers simultaneously), task arithmetic involves distilling each teacher separately and then interpolating task vectors. This reduces GPU memory usage during distillation and training complexity, as the two distillation processes are decoupled. This is proven by Table \ref{tab:ram_usage} which shows how the GPU RAM usage grows when doing ensemble distillation.
    \item \textbf{Flexibility:} Task arithmetic allows dynamic adjustment of domain contributions through interpolation weights ( $\lambda_1 = 0.9, \lambda_2 = 0.1$ in our setup). In contrast, ensemble distillation fixes the balance between domains during pre-training, requiring full retraining to adjust priorities.

\end{enumerate}

\begin{table*}[t!]
\setlength\tabcolsep{1.4pt}
\renewcommand{\arraystretch}{0.9}
\centering
\scriptsize
\caption{Model performances on speech and music tasks. All distilled models were initialized from HuBERT weights as in \cite{distilhubert}. All models were pre-trained with LibriSpeech 960, except MERT which uses Music4All dataset. * $\lambda_1 = 0.9, \lambda_2 = 0.1$ for Task Arithmetic.}
\vspace{-1.5mm}

\label{tab:main_results}
\begin{tabular}{lccccccccccccc}
\toprule
\textbf{Upstream} & \textbf{\#params} & \textbf{ASR} & \textbf{KS} & \textbf{IC} & \textbf{ER} & \textbf{SID} & \textbf{SingerID} & \textbf{PitchID} & \textbf{ESC50} & \textbf{SUPERB} & \textbf{Rank} \\
 & \textbf{(M)} & \textbf{(WER\% ↓)} & \textbf{(Acc\% ↑)} & \textbf{(Acc\% ↑)} & \textbf{(Acc\% ↑)} & \textbf{(Acc\% ↑)} & \textbf{(Acc\% ↑)} & \textbf{(Acc\% ↑)} & \textbf{(Acc\% ↑)} & \textbf{(Score ↑)} & \textbf{(Average ↓)} \\
\midrule
HuBERT (HB) & 95  & 6.42 & 96.3 & 98.34 & 64.92 & 81.42 & 81.49 & 70.04 & 72.05 & 852.52 & 1.86 \\
distil-HuBERT & 23  & 13.18 & 96.04 & 94.71 & 61.49 & 73.71 & 71.10 & 52.88 & 70.65 & 792.38 & 3.29 \\
\midrule
MERT (MR) & 95  & 23.61 & 89.45 & 59.21 & 55.19 & 29.49 & 85.69 & 91.26 & 74.00 & 699.36 & 4.43 \\
distil-MERT & 23  & 23.21 & 89.16 & 59.58 & 58.14 & 38.51 & 78.36 & 87.21 & 67.90 & 709.38	& 4.71\\
\midrule
(HB + MR) / 2 & 95 & 37.44 & 57.06 & 16.77 & 52.20 & 5.87 & 52.17 & 41.28 & 34.35 & - &  - \\
\midrule
\multicolumn{4}{l}{\textbf{Music and Speech Distilled Models}} \\
\midrule
Ensemble Distillation \cite{ensemble_distillation_speech} & 23 & 15.04 & 94.29 & 85.05 & 60.76 & 61.78 & 82.21 & 86.01 & 72.30 & 800.47 & 3.71 \\
Task Arithmetic*  & 23  & 13.97 & 95.98 & 93.59 & 62.36 & 72.20 & 83.35 & 59.99 & 70.33 & \textbf{801.78}	& \textbf{3.00}  \\
\bottomrule
\end{tabular}
\vspace{-2.5mm}
\end{table*}

\begin{table}[ht]
\centering
\scriptsize
\caption{Usage statistic comparison for distillation methods. \textbf{Task Arithmetic (case 1)} corresponds to the scenario where distillation is required before task arithmetic. \textbf{Task Arithmetic (case 2)} corresponds to the scenario in which distilled models are already available. }
\vspace{-1.5mm}
\label{tab:ram_usage}
\setlength{\tabcolsep}{1.6pt}
\renewcommand{\arraystretch}{1}
\begin{tabular}{lccc}
\toprule
\textbf{Method} & \textbf{Training Time (hrs)} & \textbf{RAM Usage} & \textbf{Params} \\
                & \textbf{(avg $\pm$ std) per epoch} & \textbf{(GB)} &  (M)\\

\midrule
distilHuBERT       & 1.10 $\pm$ 0.003  & 24 & 122 \\
distilMERT            & 1.13 $\pm$ 0.001  & 24 & 122 \\
Ensemble Distillation & 1.86 $\pm$ 0.001  & 35 & 221 \\
Task Arithmetic (case 1)        & 2.23 $\pm$ 0.003  & 24 & 122  \\
Task Arithmetic (case 2)        & 0                & 0  & --  \\
\bottomrule
\end{tabular}
\end{table}

\noindent \textbf{Question 3:} \textit{Does more pre-train data helps?}.\\

\begin{table}[t!]
\centering
\scriptsize
\caption{Performance of student models using different pre-training data. 
All rows use the same interpolation weights (0.9 for DistilHuBERT and 0.1 for DistilMERT).}
\vspace{-1mm}
\label{tab:pretrain_data_table}
\setlength{\tabcolsep}{1.9pt}
\renewcommand{\arraystretch}{0.9}
\begin{tabular}{lccccc}
\toprule
\textbf{Pre-train} & 
\textbf{ASR} & 
\textbf{SID} & 
\textbf{SingerID} & 
\textbf{PitchID} & 
\textbf{ESC50} \\
\textbf{Data} & \textbf{(WER\% ↓)} & \textbf{(Acc\% ↑)} & \textbf{(Acc\% ↑)} & \textbf{(Acc\% ↑)} & \textbf{(Acc\% ↑)} \\
\midrule
S      & 13.97 & 72.22 & 83.35 & 59.99 & 70.33 \\
M+S    & 14.01 & 72.61 & 82.00 & 60.02 & 73.63 \\
M+S+A  & 14.90 & 70.62 & 81.28 & 59.81 & 69.50 \\
\bottomrule
\end{tabular}
\vspace{-3mm}
\end{table}

\noindent Table~\ref{tab:pretrain_data_table} compares the performance of interpolated models that were pre-trained on varying combinations of data: S (LibriSpeech 960), M+S (Music4All + LibriSpeech), and M+S+A (Music4All + LibriSpeech + AudioSet \cite{audioset}). Contrary to expectations, increasing pre-training data does not consistently improve performance. While M+S achieves slight gains on SID (72.61\% vs. 72.22\%) and ESC50 (73.63\% vs. 70.33\%) compared to S, it under-performs on SingerID (82.00\% vs. 83.35\%). Furthermore, adding A (AudioSet) to the pre-training data results in performance degradation across most tasks, indicating that simply increasing the pre-training data size does not always yield better generalization.

Such results may be justified by the following reasons:

\begin{enumerate}
    \item To maintain the same number of epochs across datasets, models trained on larger data (e.g., M+S+A) required more iterations. Extended training may push the distilled models further apart in the weight space, increasing the distance between their parameter basins and degrading the effectiveness of interpolation.
    \item The distillation process may rely more on the teacher model’s ability to guide the student’s learning, which may outweigh the benefits of larger or more diverse datasets. As a result, even with more data, the student’s performance may be capped by the teacher’s representational quality.
\end{enumerate}

\noindent \textbf{Question 4:} \textit{How does interpolation weights affect performance across the tasks?}.\\

\begin{figure}[!th]
  \vspace{-4mm}
  \centering
  \includegraphics[width=1\linewidth]{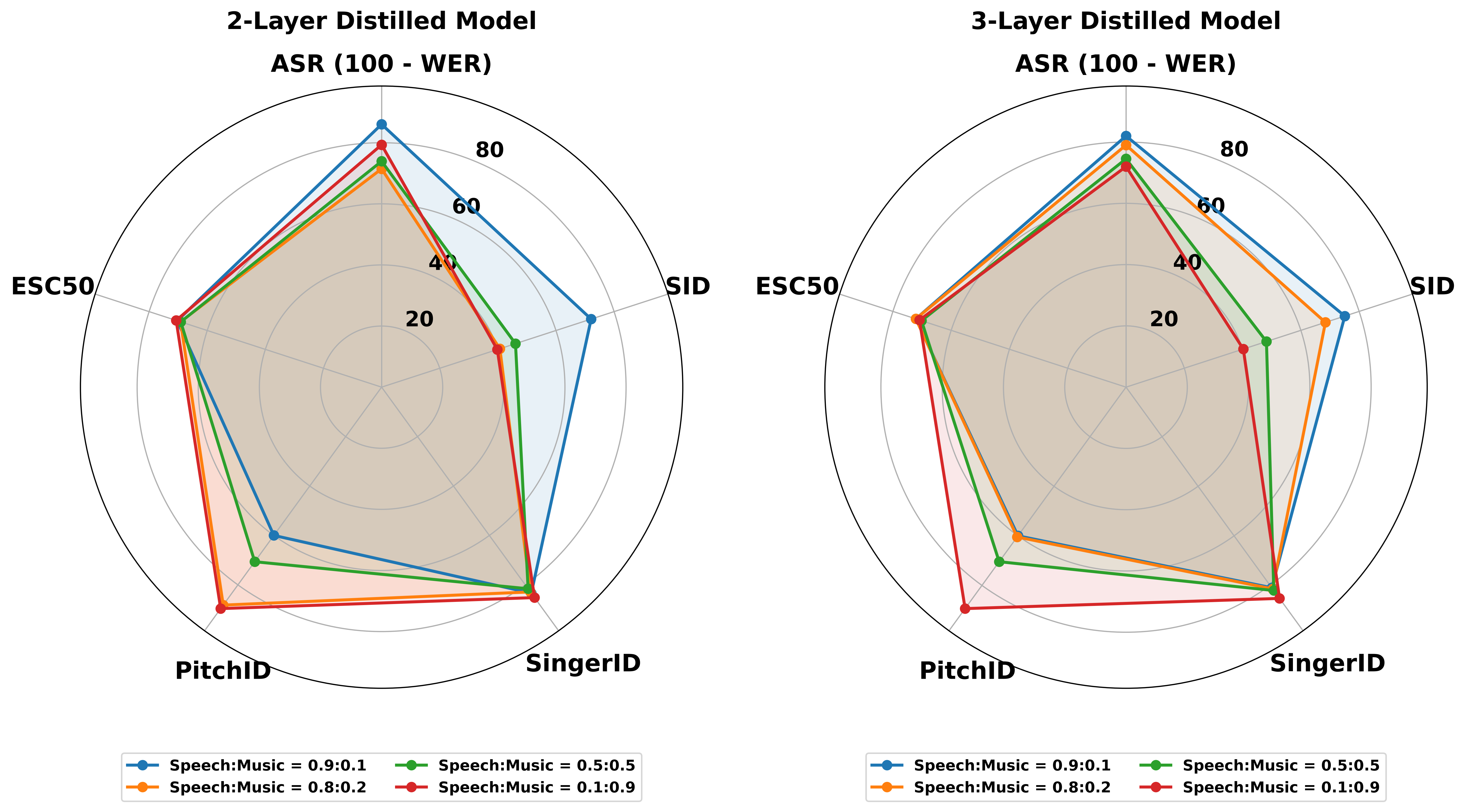}
  \caption{Performance analysis of linear interpolation of distilled models under different interpolation weights.}\label{fig:tasks_ablation_interpolation}
\end{figure}

Fig \ref{fig:tasks_ablation_interpolation} shows how the interpolation weights affect performance across speech tasks (ASR, SID), music (SingerID, PitchID), and the audio task ESC50 for interpolating distilled models with 2 (left) and 3 (right) transformer layers. From Fig \ref{fig:tasks_ablation_interpolation}, it is observed that when the interpolation weight favors speech ($\lambda_1 = 0.9, \lambda_2 = 0.1$ ), then the speech tasks perform the best with WER 13.97\% for ASR and SID of 71.10\% in a 2-layer distilled setup. Similarly, when music is favored (0.1:0.9 weight) it greatly improves the music-related tasks, with 85.15\% accuracy for SingerID and 89.60\% for PitchID. Notably, when interpolation weights are more similar e.g. 0.5:0.5, it naturally balances performance across domains yet they fail to achieve good performance in all of them. This indicates that intermediate weights may end up in a loss basin that is not close to a local minimum of either music or speech domain. Such results show that there is still space for improvement on interpolation which will be explored in future work. Finally, it can be seen that increasing student model capacity from two to three layers generally boosts performance across tasks for any given weighting. The same findings regarding model interpolation from the 2-layer distilled model are also observed for the 3-layer one. This is supported by the SUPERB score shown in Table \ref{tab:superb_compare} where the SUPERB Score is computed across all the same tasks as in Table \ref{tab:main_results}.

\begin{table}[t]
    \centering
    \scriptsize
    \caption{SUPERB scores for linear interpolation of 2-layer and 3-layer distilled models. Both at $\lambda_1 = 0.9, \lambda_2 = 0.1$.}
    \vspace{-2mm}
    \label{tab:superb_compare}
    \begin{tabular}{l c}
        \toprule
        \textbf{Distilled Model} & \textbf{SUPERB Score} \\
        \midrule
        2 layers & 801.78 \\
        3 layers & 809.32 \\
        \bottomrule
    \end{tabular}
    \vspace{-5mm}
\end{table}

\vspace{-2mm}
\section{Conclusions}

This work introduced a method for distilling multiple teacher models from distinct domains by formulating distillation as a \emph{task arithmetic} problem. We achieve this by independently distilling from a speech SSL model and a music SSL model, then computing a music and speech “task vector” for interpolation. The proposed method reduces computational overhead (GPU memory) during distillation, as there is no need to load multiple teachers simultaneously. It also mitigates distillation interference by allowing each student to learn independently. Moreover, this design supports \emph{flexible domain prioritization}: one can adjust interpolation weights to favor speech or music tasks.

While our results demonstrate that task arithmetic can successfully merge speech and music representations, there is still room for improvement. Specifically, intermediate interpolation weights appear to place the student model in suboptimal basins, as the results show degraded performance compared to strong speech or music‐biased weights. Preliminary experiments with the TIES‐merging \cite{tiesmerging} technique did not yield substantial improvements, suggesting that more sophisticated interpolation and parameter‐alignment strategies merit exploration. 


\newpage
\bibliographystyle{IEEEtran}
\bibliography{mybib}

\end{document}